\documentclass[aps,pra,preprint,longbibliography,superscriptaddress,showpacs,showkeys]{revtex4-1}
\usepackage{graphicx}
\usepackage{dcolumn}
\usepackage{bm}
\usepackage{color}
\usepackage{lmodern}
\usepackage[toc,page]{appendix}

\begin{document}

\title{Nanoscale tunnel field effect transistor based on a complex oxide lateral heterostructure}

\author{A. M\"{u}ller}
\affiliation{Institut f\"{u}r Physik, Martin Luther University Halle-Wittenberg, D-06120, Germany}

\author{C. \c{S}ahin}
\affiliation{Optical Science and Technology Center, and Department of Physics and Astronomy, University of Iowa City, IA 52242, USA}
\affiliation{Institute for Molecular Engineering, University of Chicago, Chicago, IL 60637, USA}

\author{M. Z. Minhas}
\affiliation{Institut f\"{u}r Physik, Martin Luther University Halle-Wittenberg, D-06120, Germany}

\author{B. Fuhrmann}
\affiliation{Interdisziplin\"{a}res Zentrum f\"{u}r Materialwissenschaften, Martin Luther University Halle-Wittenberg, D-06120, Germany}

\author{M. E. Flatt\'{e}}
\email{michael\_flatte@mailaps.org}
\affiliation{Optical Science and Technology Center, and Department of Physics and Astronomy, University of Iowa City, IA 52242, USA}
\affiliation{Institute for Molecular Engineering, University of Chicago, Chicago, IL 60637, USA}
\affiliation{Department of Applied Physics, Eindhoven University of Technology, Eindhoven, The Netherlands}

\author{G. Schmidt}
\email
{georg.schmidt@physik.uni-halle.de}
\affiliation{Institut f\"{u}r Physik, Martin Luther University Halle-Wittenberg, D-06120, Germany}
\affiliation{Interdisziplin\"{a}res Zentrum f\"{u}r Materialwissenschaften, Martin Luther University Halle-Wittenberg, D-06120, Germany}


\begin{abstract}

We demonstrate a tunnel field effect transistor based on a lateral heterostructure patterned from an $\mathrm{LaAlO_3/SrTiO_3}$ electron gas. Charge is injected by tunneling from the $\mathrm{LaAlO_3}$/$\mathrm{SrTiO_3}$ contacts and the current through a narrow channel of insulating   $\mathrm{SrTiO_3}$ is controlled via an electrostatic side gate. Drain-source I/V-curves have been measured at low  and elevated temperatures. The transistor shows strong electric-field and temperature-dependent behaviour with a steep sub-threshold slope 
as small as $10\:\mathrm{mV/decade}$ and a transconductance as high as $g_m\approx 22 \: \mathrm{\mu A/V}$. A fully consistent transport model for the drain-source tunneling reproduces the measured steep sub-threshold slope.
\end{abstract}


\maketitle

\section{Introduction}
Since the discovery of the electron gas between the two complex oxide band insulators $\mathrm{LaAlO_3}$ (LAO)and $\mathrm{SrTiO_3}$ (STO)\cite{Ohtomo2004} a number of devices have been realized, including both classical field transistors \cite{Foerg2012,Hosoda2013} and quantum transport devices such as the single electron transistor \cite{Cheng2011}. 
Device concepts suggested for the two-dimensional electron gas (2DEG) in LAO/STO heterostructures usually use the conducting interface in a similar fashion to how a 2DEG is used in a III-V or group-IV semiconductor heterostructure. Control of the transport occurs by electric field control of the carrier concentration at the interface, as in a field effect transistor \cite{Foerg2012,Hosoda2013}, or by gate control of the potential in a small LAO/STO island \cite{Cheng2011}. Less explored is charge transport in the STO itself, when the material is doped, for example, with vacancies or impurities \cite{Frederikse1967,Tufte1967,Lee1971}. However, transport through insulating STO over sub-micron distances is also possible when a suitable band alignment is achieved by applying electric fields.

The LAO/STO system can enable new device structures because LAO/STO islands can be used as contacts for charge injection into the STO with reliable performance and low threshold voltage. Based on such contacts we demonstrate a lateral heterostructure in which a narrow STO channel between two LAO/STO contacts conducts at bias voltages well below 100 mV, and we also demonstrate
that the current can be controlled by equally small gate-source voltages applied between a side gate and the channel. The sub-threshold slope under such conditions is very steep, indicating the importance of tunnelling currents. Therefore we present a steep sub-threshold slope device that consists entirely of oxide materials and is fabricated in a single-step, industry compatible etching process. We also demonstrate current manipulation of a wide conducting channel by a single side gate with low gate currents.

\section{Device description}
\begin{figure}
	\includegraphics[width=0.8\columnwidth]{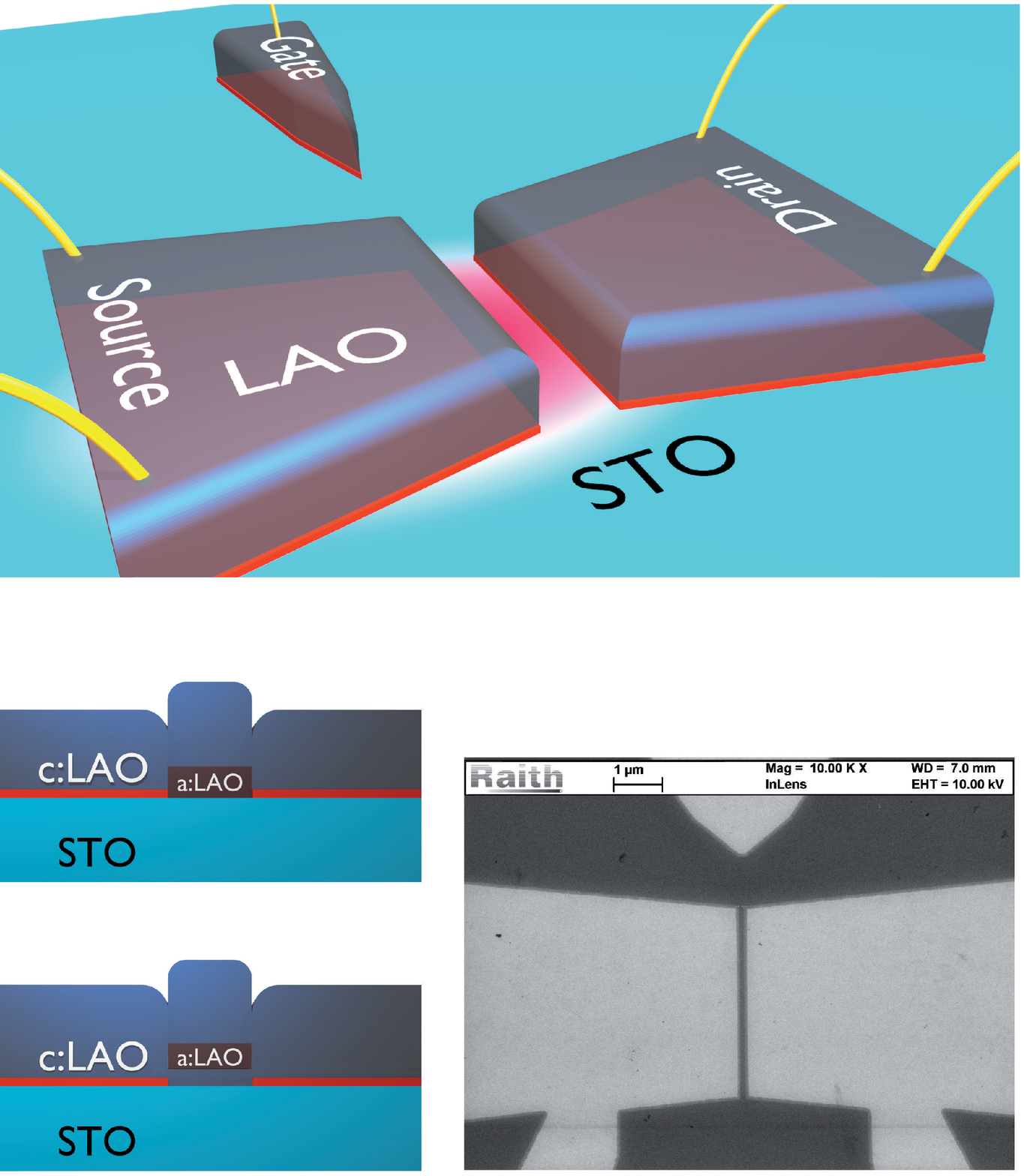}
	\caption{(a) Sketch of the sample after the structuring process. Two current pads bonded for four-probe measurements are separated electrically by a gap by removing the LAO in that region. This creates a lateral heterostructure composed of the 2DEG underneath the LAO connected laterally by the STO channel. A gate-source voltage can be applied by a side gate. The 2DEG is directly contacted by aluminium wire bonds. Alternative patterning processes use amorphous LAO (a-LAO) to avoid the formation of the 2DEG. The a-LAO is patterned by a lift off process directly on the substrate (b) or on crystalline LAO with a sub threshold thickness of two u.c. (c) as described by \cite{Schneider2006}. (d) shows an SEM picture of an etched structure.}
	\label{Structure}
\end{figure}

The device consists of an insulating STO channel which is laterally contacted by the 2DEG and a wedge-shaped side gate which is patterned into the 2DEG in the vicinity of the channel [Fig.~\ref{Structure}(a)]. The channel between source and drain has a length of $L=130-160\:\mathrm{nm}$ and a width of  $W=4\:\mathrm{\mu m}$. The spacing between the tip of the side gate and the channel is $1\:\mathrm{\mu m}$. For our measurements we use a standard lateral three terminal geometry [Fig.~\ref{Structure}(a)] consisting of two current leads (drain and source), biased by a DC-voltage source. Two additional contacts can be used as voltage probes for four terminal measurements. The gate only has a single contact. The 2DEG arises from the deposition of six unit cells (u.c.) of crystalline LAO (c-LAO) onto the $\mathrm{TiO_2}$ terminated STO surface \cite{Ohtomo2004}. The nonconducting areas are created either by removing  the c-LAO via reactive ion etching (RIE) [Fig.~\ref{Structure}(a)], as demonstrated by  Minhas et al.\cite{Minhas2016}, or by locally preventing the growth of more than three unit cells of c-LAO\cite{Schneider2006}. The latter can be achieved by using a patterned amorphous LAO layer with [Fig.~\ref{Structure}(c)] or without [Fig.~\ref{Structure}(b)] a prior deposition of a 2 u.c. subthreshold c-LAO Layer. The second process is similar to the method described by Schneider~\textit{et al.}~\cite{Schneider2006}.  The 2DEG is electronically contacted with Al wirebonds using ultrasonic bonding directly through the top LAO layer. Figure~\ref{Structure}(d) shows a scanning electron microscope picture of a typical device fabricated using the RIE process.

\section{Results}
Temperature-depended transport characteristics have been investigated in semiconducting, niobium-doped, or oxygen deficient STO\cite{Tufte1967,Frederikse1967,Lee1971,Liu2011}. In slightly reduced STO single crystals\cite{Lee1971} and reduced STO thin films\cite{Liu2011} freezeout of charge carriers was observed at low temperatures. In competition with the decrease in carrier concentration due to freezeout, the carrier mobility increases as the temperature drops and peaks at an intermediate temperature (50K for Ref.~\cite{Lee1971} and 100K for Ref.~\cite{Liu2011}).  Liu~\textit{et al.}~\cite{Liu2011} suggest a metal insulator transition occurs at this temperature, with  carrier  trapping in an oxygen vacancy donor level at lower temperatures. However, they also found the carrier trapping to be partially suppressed by an electric-field-induced detrapping. In our measurements we also observe a temperature dependent series resistance. This resistance, however, does not play a dominant role within the observed current regime for low temperatures and therefore the metal-insulator transition and detrapping effects described above are not important for interpreting our measurements.

\begin{figure}
	\includegraphics[width=1\columnwidth]{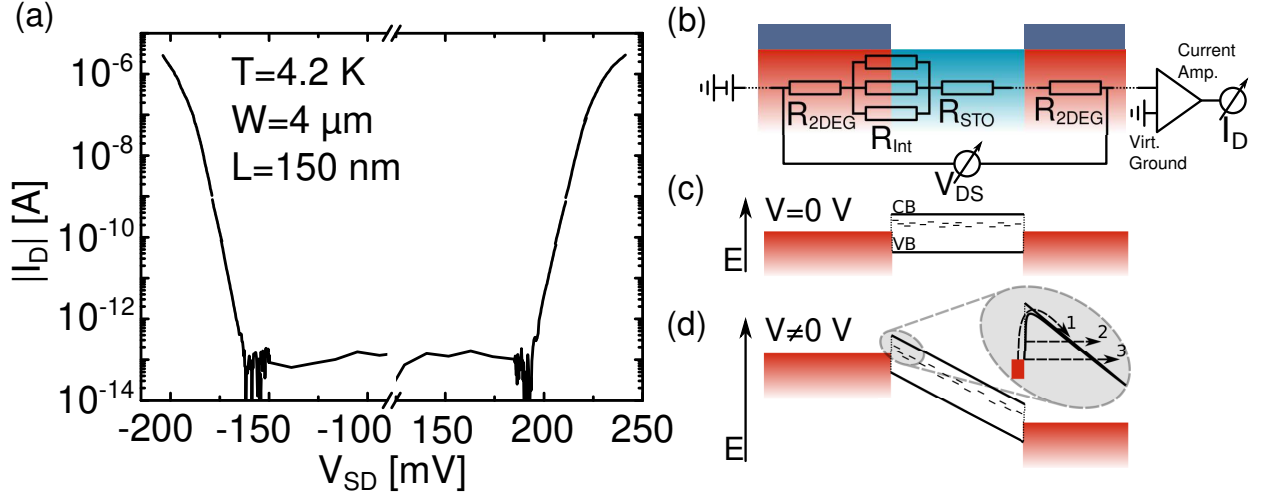}
	\caption{(a) Typical I/V-curve with a current increase of more than seven orders of magnitude measured with different current amplifier gains at $4.2\:\mathrm{K}$. This sample was patterned with the RIE process and has a nominal width of 4 $\mu$m and channel length of 160 nm. (b) Equivalent circuit shows the measurement geometry with the resistances of the 2DEG, the interface, and the channel in series. The parallel resistors at the interface represent different current injection mechanisms 1,2, and 3 shown in (d). (c) and (d) are schematic band diagrams without and with applied bias. The in-gap states represent trap sites within the STO.}
	\label{FullRange}
\end{figure}

We determine first the drain-source I/V characteristics at 4.2K of an RIE etched sample with no gate connected.  The results are shown in Fig.~\ref{FullRange}(a) and this structure will be referred to as Str$_1$ below. At low bias voltages the current is below the detection limit of the current amplifier of $\approx$100 fA. At a  threshold bias voltage, $\mathrm{V_{TH}}$, the current starts to flow and we observe more than seven orders of magnitude increase in current within a few tens of mV. This increase shows no hysteresis and is similar for positive and negative polarity (except for the current direction). The rise in current, however, is limited by an additional inherent series resistance, which, in this experiment, is of the order of a few $\mathrm{k}\Omega$ at 4.2 K leading to a constant slope dI/dV at high currents. The qualitative behavior is the same for all working structures and all  processes used, although the values for $\mathrm{V_{TH}}$ and the slope differ from sample to sample and also slightly for each cool down of the same structure. $\mathrm{V_{TH}}$ can also be shifted irreversibly to higher values by the application of a high voltage, which also results in a reduced dI/dV slope.

\begin{figure}
	\includegraphics[width=0.9\columnwidth]{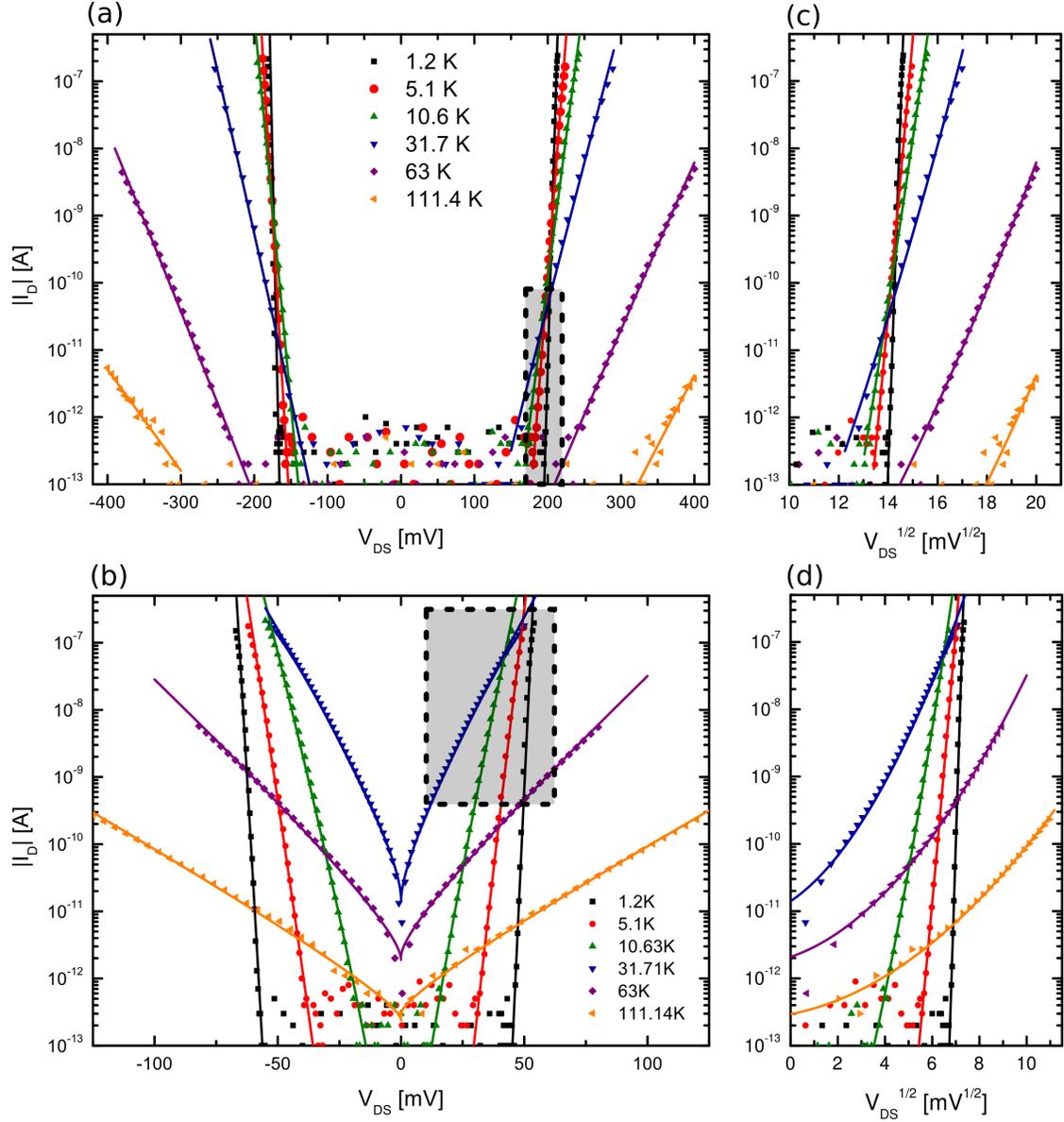}
	\caption{(a)Temperature-dependent I/V characteristics for a device with a nominal channel length of $\mathrm{L\approx160\:nm}$. The symbols represent the measured data and the solid lines represent the calculated curves. The shown measurements are in a temperature range of $1.2\:\mathrm{K} <\mathrm{T}<111.4\:\mathrm{K}$. (b) I/V-curves for a structure on the same sample with a width $\mathrm{L\approx130\:nm}$ and otherwise same dimensions.  All measurements were taken at the same run for each temperature. The inserted dashed squares cover similarly sized current and voltage ranges for each structure. In (c) and (d) the current is displayed as a function of $\sqrt{\mathrm{V_{DS}}}$}.
	\label{TempFull}
\end{figure}

In Fig.~\ref{TempFull}(a) the I/V-curves in a temperature range of $1.2\:\mathrm{K}<T<111\:\mathrm{K}$ from Str$_1$ are shown. The I/V-curves are linear in a semi logarithmic plot and their slopes decrease with increasing temperature. When the curves are described by the simple expression $I(V)=\alpha\mathrm{exp}(\beta V)$ we find $\alpha$ increases and $\beta$ decreases with rising temperature. This leads to crossing points between the curves. For $T\geq31\:\mathrm{K}$ these crossing points are below our measurement limit since $\beta$ decreases faster than $\ln\alpha$ increases. The apparent shift in $\mathrm{V_{TH}}$ is a consequence of this behavior also because $\mathrm{V_{TH}}$ is just the crossing point of the I/V-curve with our current detection limit.

For a second structure (Str$_2$) with a shorter channel ($\sim130\:\mathrm{nm}$) on the same sample [Fig.~\ref{TempFull}(b)] the behaviour is qualitatively similar but with some quantitative differences. In parts (c) and (d) of Fig. \ref{TempFull} the I/V-curves are drawn as $\mathrm{log}(I)$ vs.  $\sqrt{V}$, as is helpful when the curves are dominated by Schottky or Poole-Frenkel emission\cite{Sze1967,Mead1962,Lengyel1966} (see below). However, it is observed that the overall resistance is smaller (which may appear as a smaller $\mathrm{V_{TH}}$) and the intersections are at different current and voltage values within the measured range. In addition, for higher temperatures the I/V-curves show an increasing slope at  smaller voltages within the semi-logarithmic plot. It is noteworthy  that the linear extrapolations of the curves of Fig.~\ref{TempFull}(d) yield results very similar to Fig.~\ref{TempFull}(c) and in some cases even parts of the I/V-characteristics look almost geometrically identical [dashed squares in Figs.~\ref{TempFull}(a),(b)] even at different temperatures. This suggests that both sets of curves reflect two different parts of the same universal I/V characteristics.  Nevertheless for both Str$_1$ and Str$_2$ a slight asymmetry between positive and negative bias can be observed as the respective threshold voltages  $\pm\mathrm{V_{TH}}$ differ by up to $\Delta\mathrm{V_{TH}=25\:mV}$.

\begin{figure}
	\includegraphics[width=0.75\columnwidth]{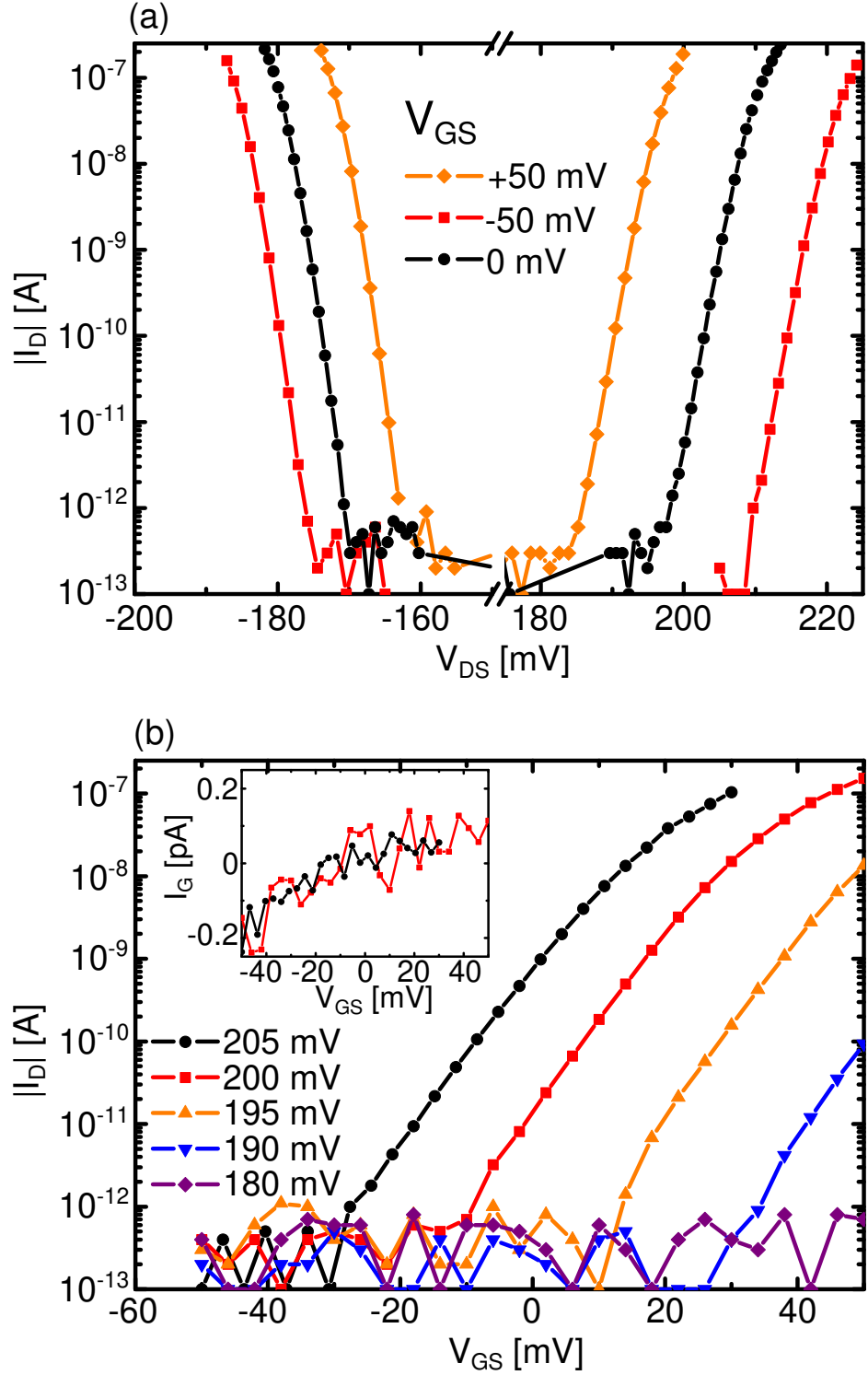}
	\caption{(a) Drain-source  voltage sweeps for $\mathrm{V_{GS}}$ of $0, +50\:\mathrm{mV}$, and $-50\:\mathrm{mV}$ at $1.2\:\mathrm{K}$. (b) $\mathrm{V_{GS}}$ sweeps for drain-source voltages in the range of $180-205\:\mathrm{mV}$. The inset shows the gate current for $\mathrm{V_{GS}}$ in the range of $\pm50\:\mathrm{mV}$.}
	\label{Gate_Dependence_2}
\end{figure}

After connecting  the side gate,  I/V-curves are taken with a fixed gate-source voltage $\mathrm{V_{GS}}$ and variable drain-source voltages $\mathrm{V_{DS}}$ [Fig.~\ref{Gate_Dependence_2}(a)], as well as with a fixed  $\mathrm{V_{DS}}$ and a variable $\mathrm{V_{GS}}$ [Fig.~\ref{Gate_Dependence_2}(b)], corresponding to transistor output and transfer characteristics, respectively. In Fig.~\ref{Gate_Dependence_2}(a) the I/V-curves of Str$_1$ taken at $T=1.2\:\mathrm{K}$  with fixed $\mathrm{V_{GS}}\in \mathrm{\{-50\:mV, 0\:mV, 50\:mV\}}$ are shown. The application of $\mathrm{V_{GS}}$ results simply in a shift of the I/V-curves to higher (lower) absolute voltages for negative (positive) values of $\mathrm{V_{GS}}$. The slope of the semi-logarithmic I/V-curves, $d(\ln I)/dV_{DS}$, remains unchanged.

\begin{figure}
	\includegraphics[width=1\columnwidth]{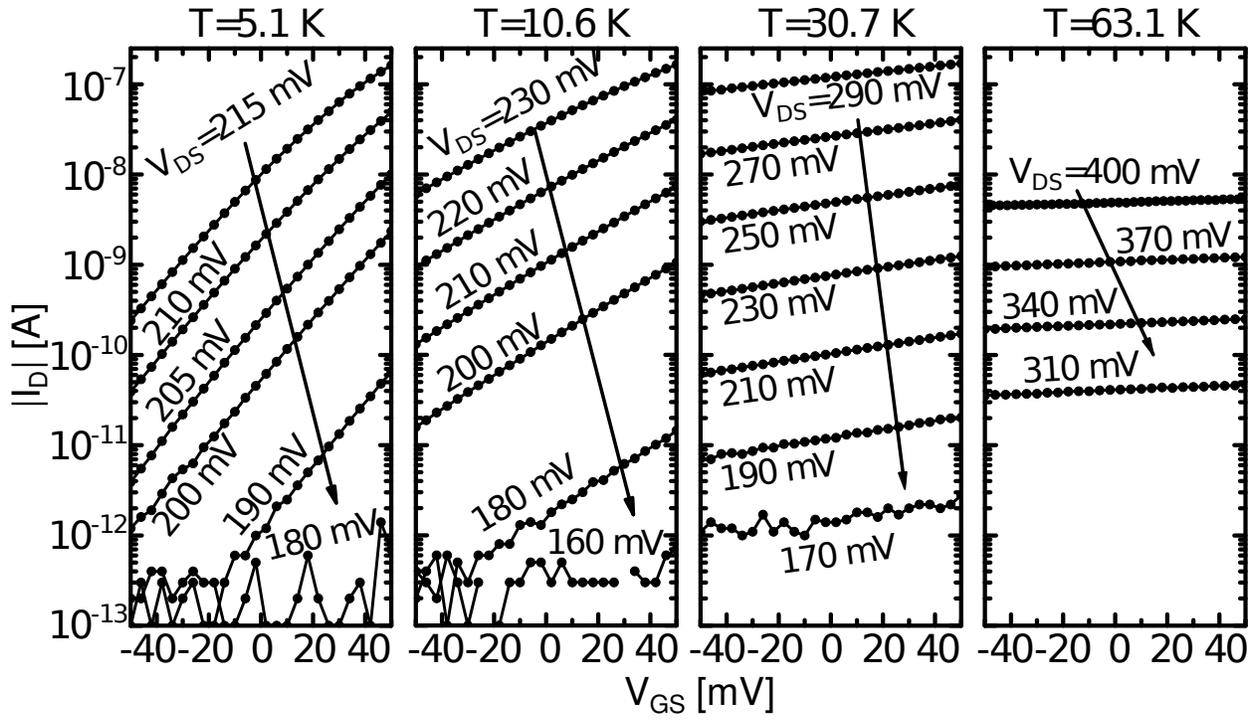}
	\caption{Temperature dependent $\mathrm{V_{GS}}$ sweeps for several $\mathrm{V_{DS}}$ taken for the same structure as shown in Fig. \ref{Gate_Dependence_2}.}
	\label{Gate_Dependence_Temp}
\end{figure}

\begin{figure}
	\includegraphics[width=1\columnwidth]{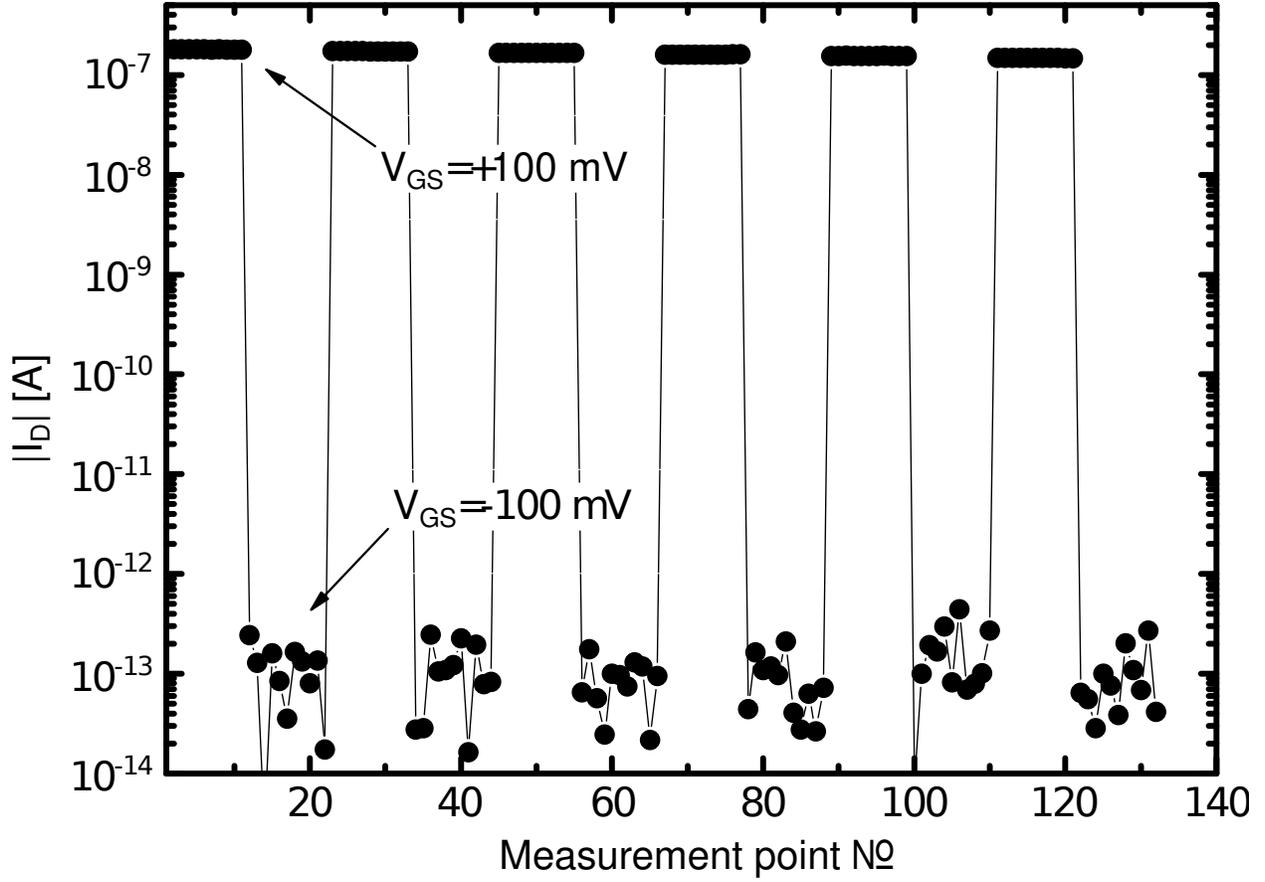}
	\caption{Demonstration of an on-off switch at 1.2 K for an applied drain-source voltage of +57 mV and $\mathrm{V_{GS}}=\pm 100$ mV. The off current is at the detection limit and the on-off ratio is at least $10^6$.}
	\label{Switching}
\end{figure}

A measurement with a $\mathrm{V_{GS}}$ sweep for different $\mathrm{V_{DS}}$ at T=1.2 K is shown in Fig.~\ref{Gate_Dependence_2}(b).  For currents $I\leq10^{-8} \:A$ an exponential dependence on $\mathrm{V_{GS}}$ is observed. For higher currents the slope starts to decrease, although the exponential behavior is evident only at lower temperatures. This fact, however, may also be attributed to the sharply reduced slope of the curves at higher temperatures which makes the influence of a series resistance less prominent. The insert in Fig. \ref{Gate_Dependence_2} (b) shows that the gate current  is well below 200 fA during the whole measurement. The gating effect is strongly reduced with increasing temperature as shown in Fig.~\ref{Gate_Dependence_Temp}. Figure~\ref{Switching} shows the reproducible operation of the device as an on-off switch. $\mathrm{V_{GS}}$ is switched between -100 mV and 100 mV repeatedly, each time switching $I_{DS}$ from $\approx 0.1\:\mathrm{pA}$ to $\approx 0.17\:\mathrm{\mu A}$, and back. The transconductance of this measurement is $g_m={\Delta I_{DS}}/{\Delta V_{GS}}\approx 0.85 \: \mathrm{{\mu A}/{V}}$ but can be as high as $22 \: \mathrm{{\mu A}/{V}}$ for higher $I_{DS}$.

In order to theoretically describe the observed behaviour one needs to consider the strong field and temperature dependence. The reduction of ${d(\ln I)/dV}$ with temperature indicates that the electric field effect scales with temperature. A higher current at low bias voltages for higher temperatures suggests a thermally-activated energy expression of the form  ${\exp({-\Phi/k_{B}T})}$. In Fig.~\ref{FullRange}(b) an equivalent circuit of the ungated structure is drawn. It describes the two areas of the 2DEG as metallic contacts connected by the insulating STO channel. This can be considered to be analogous to two back-to-back Schottky diodes with the two depletion regions merging into a single central one. For current to flow the electrons need to overcome the potential barrier at the 2DEG-STO lateral interface, which occurs via different mechanisms that can be described as electrode limited processes\cite{Simmons1971}, that together determine an interface resistance as shown in Fig.~\ref{FullRange}(c),(d). Each process can be described by a resistor, and the rate of different processes add, which corresponds to multiple interface resistors added in parallel. Because the channel's length exceeds 100~nm, transport through the bulk of the  STO, whether through the conduction band or via trap related processes, should be treated separately.

For transport through the interface, the emission of carriers from the  2DEG into the STO's conduction band
dominates the transport properties. For low temperatures and high electric fields a triangular shaped barrier is formed and charge carriers can tunnel directly into the insulator's conduction band as shown in Fig.~\ref{FullRange}(d, process 3) through Fowler-Nordheim (FN) tunneling\cite{Fowler1928,Murphy1956,Hill1967}. For higher temperatures and lower electric fields thermally activated carriers tunnel through the thinner effective potential barrier at higher carrier energies [Fig.~\ref{FullRange}(d, process 2)]. At even higher temperatures, thermionic, or Schottky, emission over the barrier takes place. The  electric field added by the image force potential lowering (Fig. \ref{FullRange} d, process 1) leads to thermionic emission\cite{Murphy1956,Hill1967}. The regimes of high and low temperature behavior always depend on the barrier height and geometry.
Murphy \textit{et al.} and Hill\cite{Murphy1956,Hill1967} also derive a temperature-dependent FN equation. In a so-called FN plot
$\left[ \mathrm{\ln(I/V^2)\:versus\:1/V}\right]$ our $I/V$-curve [Fig. \ref{FullRange}(a)] is linear, a fact which is often claimed as evidence for FN tunneling. However, the FN equation shows only an increase in current density with increasing temperature, and the slope (in an FN plot) does not exhibit the explicit temperature dependence observed in our measurements. A change in slope would occur for a change in barrier height. This, however, would not result in the crossing of the $I/V$-curves that we observe. Also it should be noted that the strong dependence of the static dielectric constant $\epsilon _{r} (T,E)$ on electric field\cite{Hegenbarth1964,Neville1972,Christen1994} is not relevant for the tunneling process because the transit time through the thin barrier is too short for the lattice to respond (as pointed out by Scott\cite{Scott1999}).

Thermionic emission with barrier lowering by the Schottky effect includes both a strong temperature and field dependence and may be described by
\begin{equation}
	J_{S}=\frac{4\pi m ek_{B}^2}{h^3}T^2 \times \mathrm{exp}\left[-\frac{\Phi_{eff}-\beta_{S}\sqrt{\cal{E}}}{k_{B}T}\right]
	\label{SchottkyEmission}
\end{equation}
with $\beta _{S}=({{e^3}/{4\pi\epsilon _0\epsilon _r}})^{1/2}$, $m$  the effective electron mass, $e$  the electron charge, $\mathrm{k_{B}}$  the Boltzmann constant, $T$  the absolute temperature, $h$ as Planck's constant, $\cal{E}$  the magnitude of the electric field, $\mathrm{\epsilon_{r}}$  the relative dielectric constant, and $\mathrm{\epsilon_0}$ the permittivity of vacuum. 
Equation~(\ref{SchottkyEmission}) leads to higher currents for higher temperatures at any given electric field. It also predicts intersection points between $I/V$-curves which move to higher current and electric field with increasing temperature. This, however, does not match our observations for all temperatures as in some cases the crossing points also move to lower current and electric field with increasing temperature as shown in Fig.~\ref{TempFull}.

However, when fitting these $I/V$-curves with~Eq. (\ref{SchottkyEmission}) one is able to calculate an effective barrier height $\Phi_{eff}$ for each point $(\mathrm{I,V/\cal{E}})$. By doing so an additional decreasing linear dependence of $\Phi_{eff}$ with respect to applied voltage emerges which leads to $\Phi_{eff}=\Phi_0+\mathrm{a V_{DS}}$. This form of $\Phi_{eff}$ in Eq. (\ref{SchottkyEmission}) accurately fits both Str$_1$ and Str$_2$, as shown in Fig.~\ref{TempFull}. The fitting parameter values are asymmetric with respect to the sign of the voltage  as discussed below.
Since for usual Schottky Emission no crossing between $I/V$-curves can occur with a fixed potential barrier $\Phi_{eff}$, the curve fits result in an increasing $\Phi_{0}$ for increasing temperature. That leads, in combination with the decreasing slope of the semi-log curve as temperature increases, to the creation of crossing points. Those points can be identified by the model via the formula
\begin{equation}
2\cdot\mathrm{\ln\left(\frac{T_1}{T_2}\right)}=\frac{a_1V+\Phi_{0,1}-\beta_{S}\sqrt{\cal{E}}}{k_BT_1}-\frac{a_2V+\Phi_{0,2}-\beta_{S}\sqrt{\cal{E}}}{k_BT_2}
\end{equation}
using the condition $\mathrm{J_1(T_1)=J_2(T_2)}$ and solving for $V$ by taking $\mathrm{\cal{E}=V/L}$.
As a result, $I/V$-curves taken at different temperatures can show crossing points.

In our picture of two back-to-back Schottky diodes one contact is always biased in reverse and the other in forward direction. Due to its higher resistance only the reverse biased contact needs to be considered. Within that picture small differences in the barrier height at the two different sides lead to an asymmetry of the I/V-curves that must vanish at higher temperatures because of the decreasing $\Phi/(k_{B}T)$. The height of such a barrier may be lowered locally when $\mathrm{O^{2+}}$ vacancies are present at the interface\cite{Numata2006}. In consequence, a different respective spatial distribution of vacancies at the two interfaces results in an asymmetry with respect to the sign of the bias voltage. This distribution of vacancies may be altered by the application of a high bias voltage leading to irreversible changes of the I/V-curves (see Supplemental Material for shifted $I/V$-curves after high bias voltage). The irreversibility results from the fact that vacancies pushed into the 2DEG cannot travel back. These vacancies are exposed to a much smaller electric field within the 2DEG than in STO. Therefore, we only observe an increase in $\mathrm{V_{TH}}$ and never a decrease, because the vacancy concentration inside the channel can only decrease.
In addition to the $\mathrm{O^{2+}}$ vacancies, surface and interface states are expected to exist. Due to possible variations in the densities of the interface states, the work functions are expected to be different at the two separate junctions \cite{Sze2006}.

In addition to current flow via the conduction band, Lee \textit{et al.}~\cite{Lee1971} suggested the existence of an impurity band in slightly oxygen reduced STO with strong temperature dependent properties, due to the combination of low doping and the large temperature dependent static dielectric constant\cite{Hegenbarth1964,Neville1972,Christen1994}. Increasing the electric field and temperature reduces the static dielectric constant and thus increases the potential between the vacancies. Carriers trapped at the vacancies can surpass this potential more easily at elevated temperatures. This behaviour would also indicate that the potential barrier into such an impurity band should be highly dependent on temperature and increase with increasing temperature.

By the application of $\mathrm{V_{GS}}$, the bands within the STO are shifted up or down depending on $\mathrm{V_{GS}}$'s sign. This shift is expressed in a rise or reduction in $\mathrm{\Phi_{0}}$ through $\mathrm{\Phi_{0}\pm\alpha_{GS}\times V_{GS}}$ which results in an exponential change in current when sweeping $\mathrm{V_{GS}}$. Since $\mathrm{\pm\alpha_{GS}\times V_{GS}}$ is also divided by $\mathrm{k_{B}T}$ a strong decrease in the gating effect is observed with increasing temperature.

As Fig. \ref{TempFull} shows the model fits the measurements almost perfectly; the lines indicate the fitted curves and the symbols represent the measurements. The crossing points are now linked to a change in effective barrier height $\mathrm{\Phi_{0}}$ and the parameter a. The parameters given by the fitting procedure are shown in Table~\ref{TableParam}. For the calculation of $\mathrm{I_D}$ an effective mass of $3\mathrm{m_e}$,
a high frequency dielectric constant of $\epsilon_r = 5$\cite{Neville1972} and an emitter area of $A=5\:\mathrm{nm}\times4\:\mathrm{\mu m}$ have been used. For reduced STO thin films Liu et al. \cite{Liu2011} determined an activation energy of the oxygen vacancies of 25 meV between 200 and 300 K. In the case of unannealed c-LAO/STO samples, Ref. \cite{Liu2013} showed an oxygen vacancy activation energy of 4.2 meV below 100 K. Both are determined by Hall measurements. This gives an additional hint that the injection mechanism is electrode limited and not bulk limited due, \textit{e.g.}, to Poole-Frenkel emission via oxygen vacancies. The increase in $\Phi_{0}$ is noticeable from the data, because at low fields one would always expect an increase in current, if the potential barrier is the same for all temperatures. Due to the different voltage regimes for the $I/V$ curves for the different structures in Fig. \ref{TempFull}, the fitting parameters have necessarily different values. However, when looking at the qualitative dependence from $\mathrm{\Phi_{0}(T)/a(T)}$ over temperature a very similar behavior for both structures can be observed. That mathematically supports our earlier statement that both structures showed similar features in their $I/V$ curves (indicated by the shaded areas in Fig. \ref{TempFull}) but at different voltages. Also, the irreversible modification of the $I/V$ characteristics can now be linked to motion of oxygen vacancies as observed in a recent study \cite{Wu2013}.

\begin{table}
	
	\caption{Parameters used for the plotting in Fig. \ref{TempFull}.}
	\label{TableParam}
	\begin{tabular}{cccccc}
		\\ \hline \hline
		Str$_1$ & T (K) & $\Phi_{0}$ (meV) & a (e)\\ \hline
		+V/-V & 1.2 & 28.19/30.12  & -0.041/0.067 \\
		+V/-V & 5.1 & 42.58/44.72 & -0.097/0.135 \\
		+V/-V & 10.6 & 55.6/59.4 & -0.132/0.192 \\
		+V/-V & 31.7 & 102.3/103 & -0.222/0.259 \\
		+V/-V & 63 & 195/199.5 & -0.275/0.301 \\
		+V/-V & 111 & 382.3/327.6 & -0.429/0.297 \\ \hline \hline
	\end{tabular}
	\begin{tabular}{cccccc}
		\\ \hline \hline
		Str$_2$ & T (K) & $\Phi_{0}$ (meV) & a (e)\\ \hline
		+V/-V & 1.2 & 15.1/15.7  & -0.082/0.056\\
		+V/-V & 5.1 & 21.4/21.4  & -0.202/0.145\\
		+V/-V & 10.6 & 25.2/25.2 & -0.265/0.207\\
		+V/-V & 31.7 & 42.2/42 & -0.322/0.296\\
		+V/-V & 63 & 101.6/102.1 & -0.374/0.373\\
		+V/-V & 111 & 209.4/211.4 & -0.403/0.414  &\\ \hline \hline
	\end{tabular}
\end{table}

Cen et al. \cite{Cen2010} showed results for smaller three-terminal structures with gating; the three terminals were conducting lines induced in an insulating LAO/STO bilayer using a conducting AFM, and the spacings between source, drain and gate are up to one order of magnitude smaller than our channel. At high temperatures they observed an increase in conductance caused by thermal activation, and they suggested quantum field emission as a dominant transport mechanism at low temperatures, supported by the signature of STO phase transitions related to changes in $\epsilon_{r}$. We suggest that the process in Ref. \cite{Cen2010} is not suitable for huge throughput and long device stability, in contrast to the RIE etching process used in our case. Due to smaller dimensions and higher applied voltages the electric field in their experiment is much higher than described here leading to a different transport mechanism with a different temperature dependence. The larger dimensions of our device simply exclude any direct tunneling process between the contacts. The difference in functionality is also visible due the fact that we observe no influence of the structural phase transitions as described in Ref. \cite{Cen2010}.

\section{Conclusion}
We have shown that it is possible to create a new type of field effect transistor based on transport through more than 100 nm of STO in an LAO/STO heterostructure. The transport is dominated by the Schottky barriers between the electron gas on both sides of the gap and the STO inside the gap resulting in strong temperature dependent and non-linear I/V characteristics. Because of the large dielectric constant of the STO the barrier height can be controlled by a side $\mathrm{V_{GS}}$ resulting in full transistor functionality. The device is fabricated using state of the art lithography and dry etching processes. Results based on two different patterning processes (see Supplemental Material at [...] for transport characteristics in amorphous-LAO/c-LAO structure) exclude transport through defects induced by dry etching. Our results show that due to the special properties of STO nanostructures in LAO/STO, there may still be an unrecognised potential for applications beyond classical device concepts. Even though the  fabricated transistor shows little effect at elevated temperatures, the design concept demonstrates a natural way to include high-k dielectric materials into a transistor by using them for the gate and the channel as well, which is an advantage over  commonly used silicon technology. Steep drain-source I/V-curves enable gating with very low voltages and lead to a very low power switching. On the other hand, the experiments show the limitations of nanopatterning of LAO/STO devices. No matter whether the gap is fabricated by etching or by the method by Schneider et al. \cite{Schneider2006}, a gap of less than 200 nm width between two regions of LAO/STO becomes conducting at relatively low bias voltages seriously limiting the density of integrated nanodevices. For integration purposes, it may be necessary to add additional 'dummy' gates between different devices in order to efficiently insulate them from each other.

\section{Acknowledgements}
This work was supported by the European Commission in the project IFOX under grant agreement NMP3-LA-2010-246102 and by the DFG in the SFB 762. We thank H.H. Blaschek for technical assistance.
M.E.F and C.\c{S}. acknowledge support for theoretical calculations and modeling from the Center for Emergent Materials,
an NSF MRSEC under Award No. DMR-1420451.

\section{Author contributions}
A. M\"uller did part of the processing, analyzed the data,  performed all the transport measurements and fit them to the model equations. C. \c{S}ahin and M. E. Flatt\'e suggested the electric-field-dependent tunneling rate and derived the resulting equations. M.Z. Minhas deposited the LAO layers by PLD. B. Fuhrmann did the reactive ion etching. G. Schmidt planned and supervised the experiment. All authors contributed to the manuscript and reviewed it prior to submission.

\section{Additional information}
The authors declare no competing financial interests.

\begin{appendices}
\section{Sample preparation}
The devices were fabricated from large area LAO/STO heterostructures. These heterostructures were fabricated by pulsed laser deposition (PLD) of 6 u.c. LAO on (001) oriented STO substrates. The substrates were prepared as described in previous studies \cite{Koster1998,Kawasaki1994}. For the deposition (fluency of $\mathrm{2\:J/cm^{2}}$ at $\mathrm{f=2\:Hz}$) a $\mathrm{O_{2}}$ pressure of 0.001 mbar at a temperature of $\mathrm{T=850^{\circ} C}$ was used.

The layers were patterned by electron beam lithography and dry etching. PMMA was used as an electron beam resist and subsequent etch mask.The exposure was done at an acceleration voltage of 30 kV using a RAITH pioneer exposure tool. After development the LAO was patterned by dry etching down to the STO substrate using the etching process described in \cite{Minhas2016}. The a-LAO layer was patterned by a standard PMMA lift off process. For the process described in Fig. \ref{Structure} the a-LAO layer was annealed for 1 h at $\mathrm{650^\circ C}$ in $\mathrm{O_2}$ atmosphere in order to make the interface insulating \cite{Liu2013}. No contact metallization is used but the electron gas is contacted electrically by direct ultrasonic bonding through the LAO.

\section{Measurement}
The samples were characterized in a $^4\mathrm{He}$ bath cryostat with a variable temperature insert which allows measurements down to 1.2 K. Voltages were applied using high precision home built 20 bit digital to analog converters. The source current was measured either by a multiple range current amplifier with a noise floor of approximately 200 fA or a home build current amplifier with fixed gain. The gate current was measured via the voltage drop over a $10\: \mathrm{M\Omega}$ series resistor. All voltages were measured using high precision high impedance difference amplifiers connected to an Agilent 34420A nanovoltmeter.
\end{appendices}

\clearpage
\clearpage

\clearpage

\clearpage

\clearpage

\clearpage

\clearpage
\end{document}


\section{Supplemental Information}

\subsection{Prestructured substrate}
\begin{figure}
	\includegraphics[width=0.80\columnwidth]{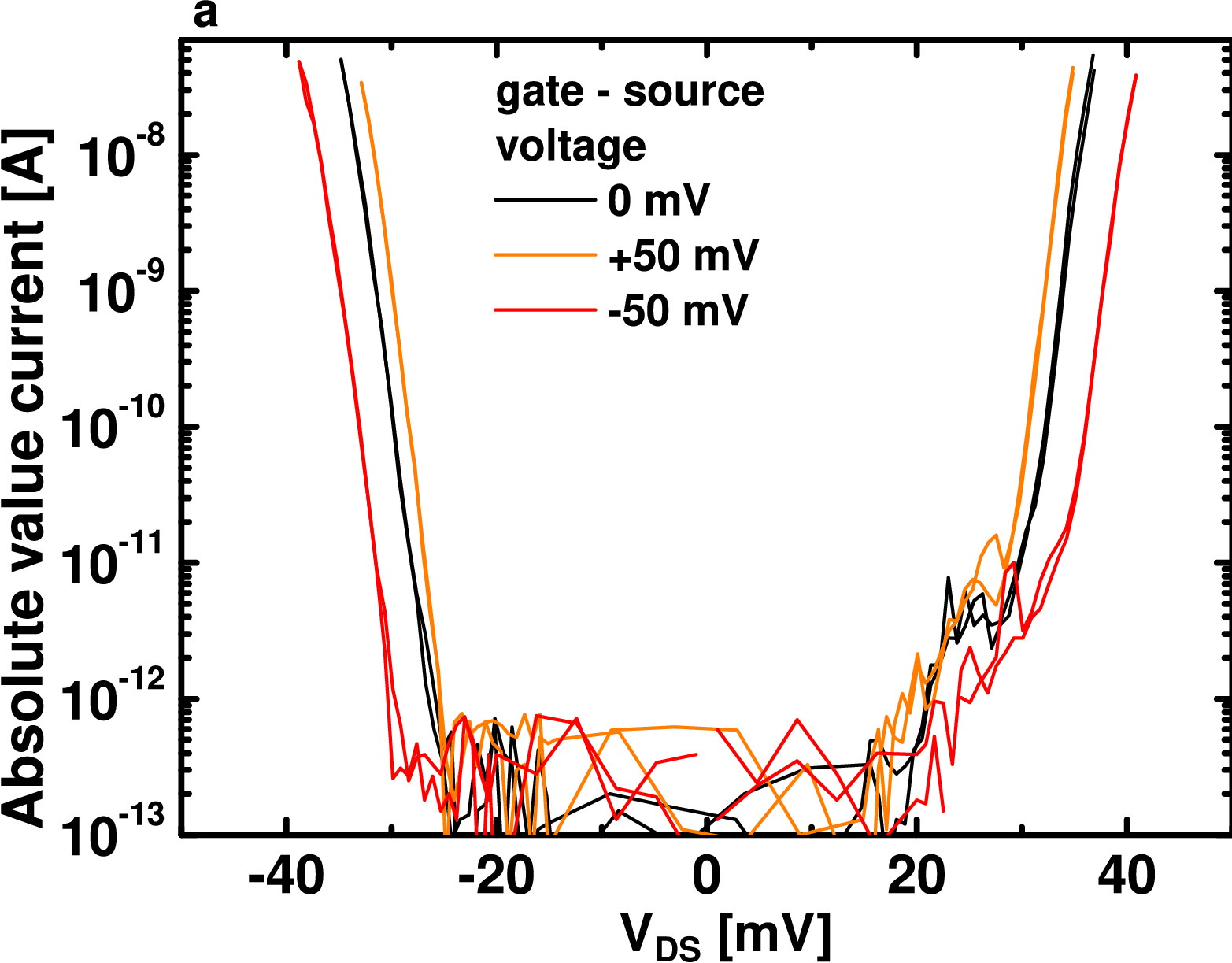}
	\includegraphics[width=0.80\columnwidth]{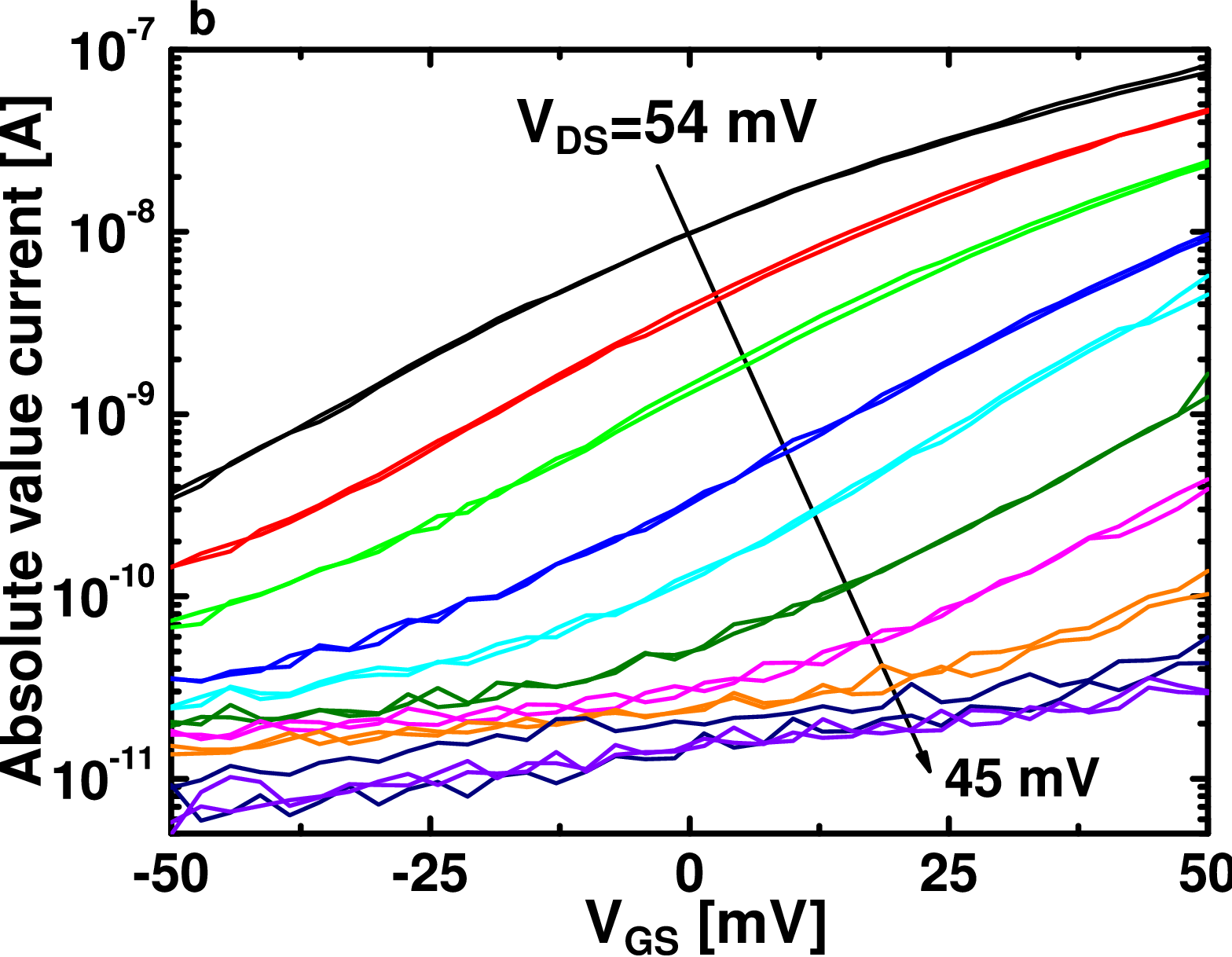}
	\caption{In (a) drain-source voltage sweeps at different gate-source voltages at 1.2 K. (b) displays gate-source sweeps for fixed $\mathrm{V_{SD}}$ at 1.2 K.}
	\label{2700_amo_LAO}
\end{figure}

In Fig. \ref{2700_amo_LAO} gated I/V-curves, taken at 1.2 K, are shown similar to the measurements displayed in Fig. 3 (Manuscript), 4 (Manuscript). The sample used here (Str$_3$) are processed via prestructureing the substrate with an amorphous LAO layer with subsequent annealing. The I/V-curve Fig. \ref{2700_amo_LAO} (a) shows an additional shoulder resulting also in an reduced slope within the gate-source sweeps taken near that position shown in Fig. \ref{2700_amo_LAO} (b). Also the backsweep is shown in both figures which shows no hysteresis.
\clearpage
\subsection{Threshold shifting}
\begin{figure}
	\includegraphics[width=0.80\columnwidth]{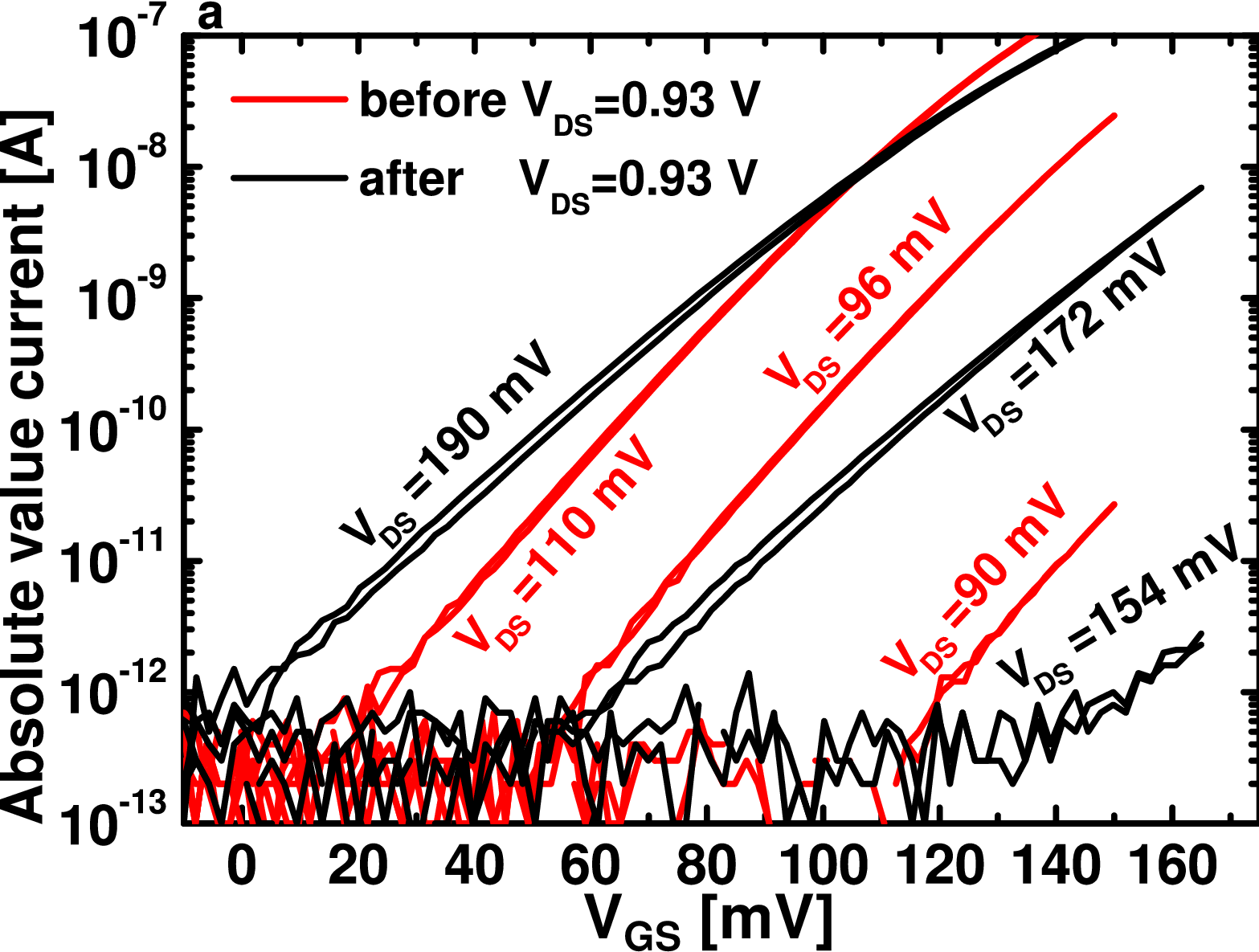}
	\includegraphics[width=0.80\columnwidth]{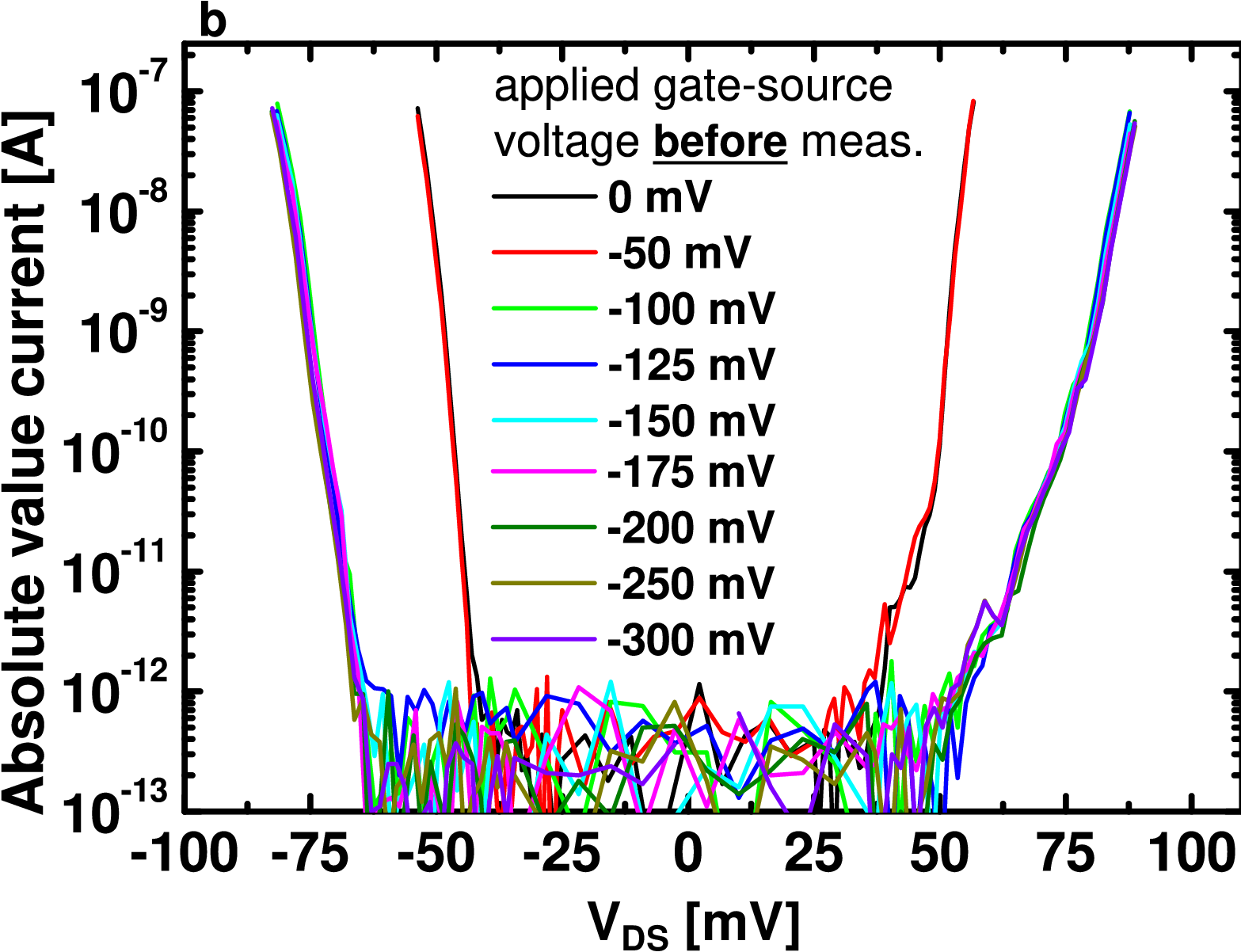}
	\caption{Measurements taken at 1.2 K. In (a) gate-source sweeps are shown before and after the application of an hight drain-source voltage of 0.93 V (Str$_1$). In (b) drain-source sweeps are drawn taken after the application of different gate-source voltages (Str$_3$).}
	\label{Shift}
\end{figure}

As mentioned above the I/V-curves can be shifted by the application of high voltages. That shift can be induced by $\mathrm{V_{GS}}$ and $\mathrm{V_{DS}}$ an lead to reproducible results afterwards as shown in Fig. \ref{Shift}. By an application of a high voltage we observed always an increase in resistance (higher $\mathrm{V_{TH}}$) independent of the voltage sign. Also should be noted, that $\mathrm{V_{TH}}$ is changed on both sides. The exponential slope $\mathrm{dI/dV_{GS}}$ reduces strongly with the shift as seen in Fig. \ref{Shift} (a).
